\documentclass[12pt,a4paper]{article}
\usepackage{a4wide}
\usepackage{amsmath}
\usepackage{amsfonts}
\usepackage{cite}
\usepackage{graphicx}
\begin{document}

\title{Identifying non-k-separability of a class of N-qubit complete graph states using correlation tensors} 
\author{{N. Ananth and  M. Senthilvelan }\\
{\small Centre for Nonlinear Dynamics, School of Physics, Bharathidasan University,} \\{\small Tiruchirappalli - 620024, Tamilnadu, India} }
\date{}
\maketitle

%\section*{Abstract}
\begin{abstract}
We derive a general expression for standard tensor norm of $N$-body correlation tensors 
for $N$-qubit complete graph states. 
With the help of this expression, we formulate a separability criterion that identifies non-$k$-separability of 
a class of $N$-qubit complete graph states, including GHZ state.  
We illustrate the performance of our criterion by considering the $N$-qubit complete graph states   
added with colored product noise. 
We also demonstrate that few local measurement settings are sufficient to evaluate our criterion. 
\end{abstract}

\section{Introduction}
\label{intro}
Multipartite entanglement plays a prominent role in 
quantum error correction \cite{schl2002}, one-way quantum computer \cite{raus2001,raus2003} and 
quantum secret sharing \cite{hill1999}. 
The genuine multipartite entangled states eventhough act as great resource for various quantum information and 
computational tasks \cite{horo2009,guhne2009},  
they become partially entangled states or mixed states due to decoherence. 
These partially entangled states are also utilized in certain quantum information processes, see for example 
\cite{li2000,rigo2009,wang2009,gord2010,na2015}. 
To certify the multipartite entangled states,   
several entanglement identifiers in the form of inequalities  
\cite{seev2008,hube2010,gao2013,guhne2010,gitt2010,asch2004,badz2008,hasan2008,lask2011,julio2011,anan2015a,anan2015b} 
have been constructed. 
However, for an experimental implementation, the entanglement conditions should be 
represented in terms of local observables \cite{seev2008,hube2010,gao2013}.   
One of the widely studied experimentally accessible entanglement identifiers is  
the entanglement witness \cite{bour2004,kram2009}. 
Recently two different tools, namely correlation tensors and covariance matrices approach,  
have been introduced as an alternate to the entanglement witnesses \cite{gitt2010,julio2011}. 
Among these two, the correlation tensors approach got wide attention since it accounts 
all $N$-body correlations that contains all information about the entanglement of a system.   
However, recent results reveal that 
single particle information can also be used to 
distinguish several important classes of entanglement by associating it with a geometric object, 
namely entanglement polytope \cite{walt2013}. 
With the help of correlation matrix approach several entanglement conditions 
for bipartite and multipartite states were reported 
\cite{asch2004,badz2008,hasan2008,lask2011,julio2011,julio2007,julio2008}.    
For example, the experimentally friendly conditions for entanglement were derived 
in terms of correlation function, 
which are stronger than the entanglement witnesses \cite{badz2008}.
Later a geometric approach has been developed in order to identify the  
genuine multipartite entanglement (GME) 
and non-$k$-separability of different families of multipartite entangled states \cite{lask2011}. 
Vicente and Huber have developed a general framework to detect GME and non-full separability 
in multipartite states  \cite{julio2011}.
Even though different upper bounds were derived to identify GME of 
$3$-partite and $4$-partite states \cite{julio2011},  
it is difficult to generalize them for $N$-number of systems $(N\geq 5)$. 
This observation motivates us to consider a class of multipartite entangled states and 
to characterize their entanglement using correlation tensors. 
The restriction on a class of multipartite entangled states and 
deriving a separability condition for that class offers an advantage to the study of 
non-$k$-separability of multipartite states. 

To begin, we recall the definition of $k$-separability. 
An $N$-partite pure quantum state 
$|\psi_{k\textrm{-prod}}\rangle$ is $k$-product 
$(k=2,3,\ldots,N)$ if and only if it can be written as a product of $k$ substates, 
$|\psi_{k\textrm{-prod}}\rangle = |\psi_1\rangle \otimes |\psi_2\rangle \otimes \ldots \otimes|\psi_k\rangle$,    
where $|\psi_i\rangle$, $i=1,2,\ldots,k$, represents the state of a subsystem or 
a group of subsystems \cite{gabr2010}.   
A mixed state $\rho_{k\textrm{-sep}}$ is called $k$-separable, if it can be decomposed into pure $k$-product  states, that is 
\begin{align}
\label{k2} \rho_{k\textrm{-sep}} = \sum_i p_i~ \rho_{k\textrm{-prod}}^i,  
\end{align}
where $p_i > 0$ with $\sum_i p_i = 1$. 
An $N$-partite state is called non-$k$-separable if it is not $k$-separable \cite{gabr2010}. 

We note here that the criterion based on two particle correlations 
like covariance matrix criterion is inadequate 
to detect the entanglement of graph states \cite{gitt2010}. 
Under this circumstance, we believe that 
the correlation tensor approach is fruitful  
to characterize the graph states \cite{julio2011}. 
Graph states are pure multiqubit states, essential for quantum error correction \cite{schl2002} and 
one-way quantum computation \cite{raus2001}. 
The graph states and their special types such as GHZ states and cluster states have been 
experimentally created and analysed \cite{bodi2006,leib2005,walt2005,lu2007}. 
Several attempts have been made to identify and quantify the entanglement of graph states 
\cite{hein2004,hein2005,toth2005,jung2011,guhn2011,mark2007}. 
Recently, entanglement witnesses were also constructed for graph states 
but they were limited to few qubits only \cite{jung2011,guhn2011}. 
Our survey on this topic reveals that the non-$k$-separability of graph states is 
yet to be formulated.  

Motivated by these observations, in this paper, 
we find standard tensor norm of $N$-body correlation tensors for 
different multipartite entangled states. 
The norms of the $N$-qubit complete graph states yield a suitable upper bound for 
$k$-separable $N$-qubit complete graph states.  
Using this norm we derive a separability condition that identifies the non-$k$-separability of a class of complete graph states. 
We illustrate the performance of our criterion for $N$-qubit complete graph states added with colored product noise. 
The criterion presented in this paper can be experimentally feasible with the help of 
few local observables. 

The structure of this paper is organized as follows. 
In Section \ref{sec2}, we recall correlation tensors 
and derive the norm of $N$-body correlation tensors for different multipartite entangled states. 
In Section \ref{sec3}, we formulate a separability criterion to identify the non-$k$-separability   
of a class of complete graph states, including GHZ state. 
In Section \ref{sec4}, using our criterion, we analyze the GME and non-$k$-separability 
of $N$-qubit complete graph states added with colored product noise. 
In Section \ref{sec5}, we present our conclusion and discuss the experimental feasibility of our criterion. 
We present the method of finding the norm of $N$-body correlation tensors for $N$-qubit complete graph states 
in the Appendix. 

%%%%%%%%%%%%%%%%%%%%%%%%%%%%%%%%%%%%%%%%%%%%%%%%%%%%%%%%%%%%%%%%%%%%%%%%%%%%%%%%%%%%%%%%%%%%%%%%%%%%%%%%%%%%%%%%%%%%%%%%%%%%%%%%%%%%%%%%%%%%%%%%%%%%%%%%
\section{Correlation tensors}
\label{sec2}
To begin, we recall correlation tensors. 
We then briefly review complete graph states and present the standard tensor norm of 
correlation tensors of certain genuine multipartite entangled states.  

The Bloch representation of an $N$-qubit density operator is represented by \cite{bloc1946,hioe1981}
\begin{align}
\rho = \frac{1}{2^N} \bigotimes_{n=1}^N \left( \mathcal{I}_2^{(n)} + \sum_{i_n} t_{i_n}^{(n)}~\sigma_{i_n}^{(n)}\right). 
\label{blonqdm}
\end{align}
Here $\mathcal{I}_2^{(n)}$ stands for the identity operator,  
$\sigma_{i_n}^{(n)}$ are Pauli matrices and $t_{i_n}^{(n)}=\langle \sigma_{i_n}^{(n)}\rangle$,  
$i_n=1,2,3$ and $n=1,2,\ldots,N$. 
For our convenience, after making the tensor product in (\ref{blonqdm}), 
we redenote the coefficients $t_{i_1}t_{i_2}\ldots t_{i_N}$ as $t_{i_1,i_2,\ldots, i_N}$. 
It is clear from the above that the correlation tensors can be represented by expectation values of  
all possible tensor products of Pauli operators \cite{julio2011,julio2007,julio2008}. 
The $N$-body correlation tensor or full correlation tensor,  
denoted by $\mathcal{T}_{i_1,\ldots,i_N}$, is defined by \cite{julio2011} 
\begin{align}
\mathcal{T}_{i_1,\ldots,i_N} = \langle \sigma_{i_1}^{(1)}\otimes \sigma_{i_2}^{(2)} \otimes \ldots \otimes \sigma_{i_N}^{(N)} \rangle.  \label{ct}
\end{align}
We consider the standard tensor norm of $N$-body correlation tensors as given in \cite{julio2011}, that is   
\begin{align}
||T_{i_1,i_2,\ldots,i_N}|| = \sqrt{\sum_{i_1,\ldots,i_N} \mathcal{T}_{i_1,\ldots,i_N}^2} \label{st}. 
\end{align}
Since this norm is multiplicative under outer products, 
we can factorize the $N$-body correlation tensors into product of 
$k$ number of lower order correlation tensors as an upper bound 
to derive the separability criterion. 

\begin{figure}[h]
\begin{center}
\resizebox{.5\columnwidth}{!}{\includegraphics{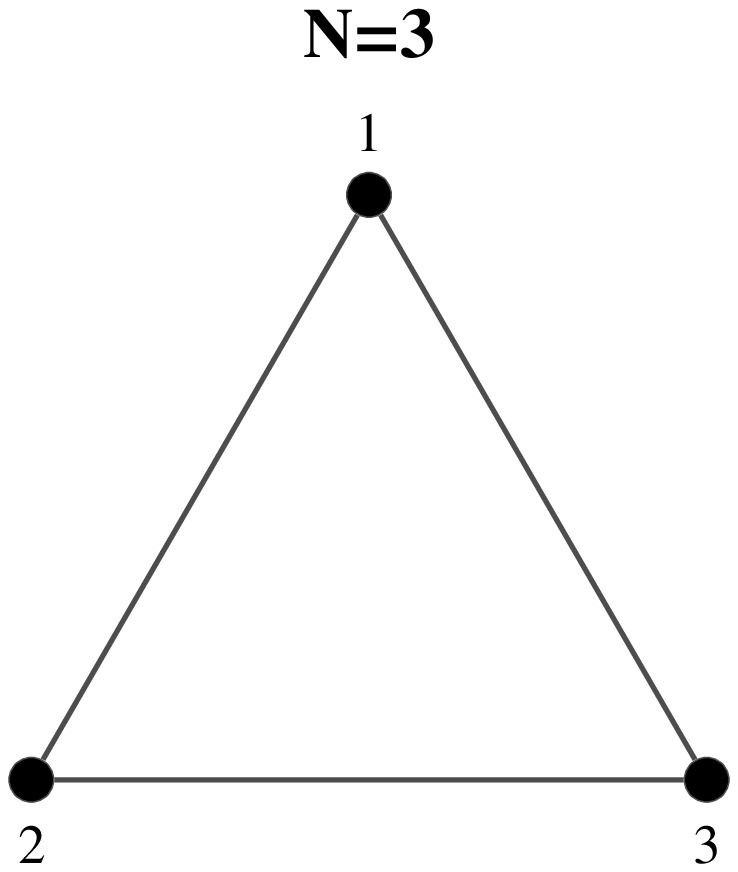}\quad \includegraphics{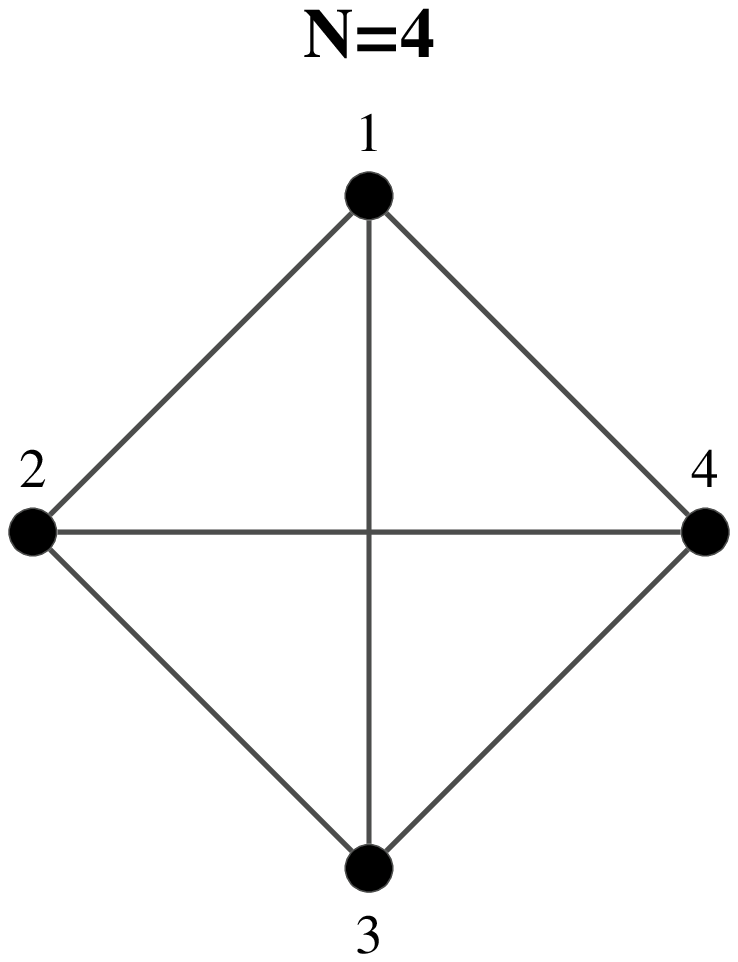} \quad \includegraphics{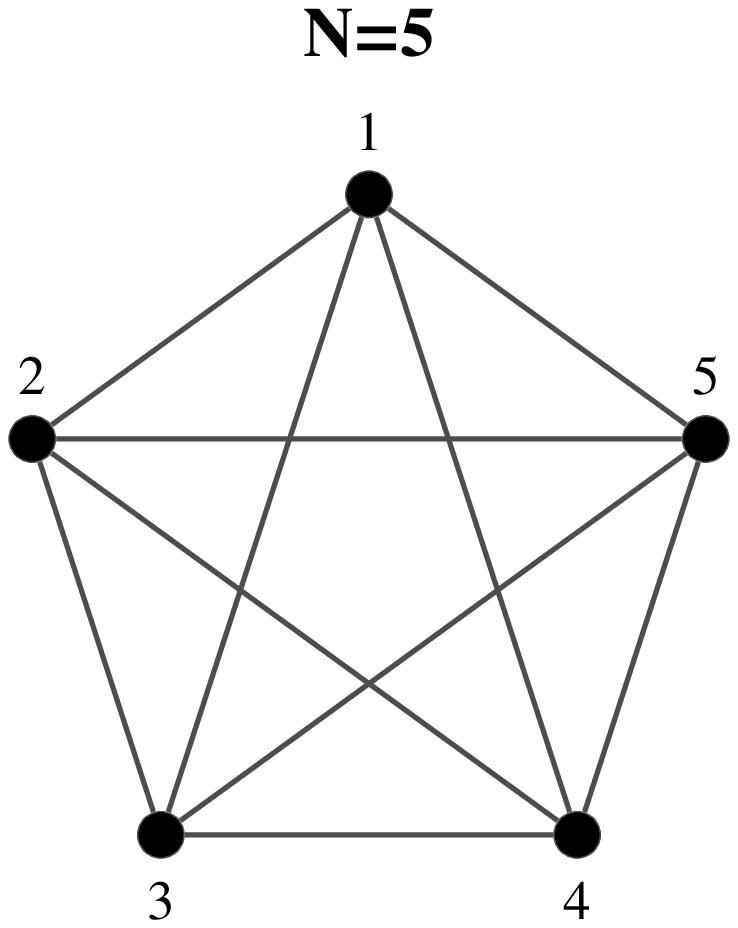}} 
\resizebox{.5\columnwidth}{!}{\includegraphics{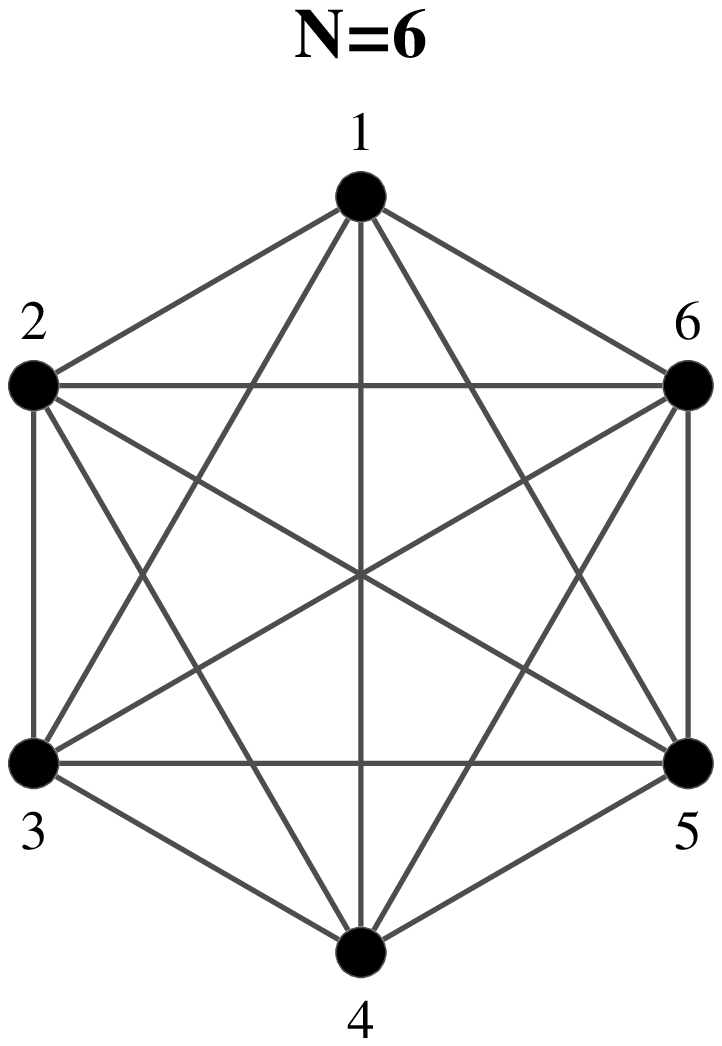}\quad \includegraphics{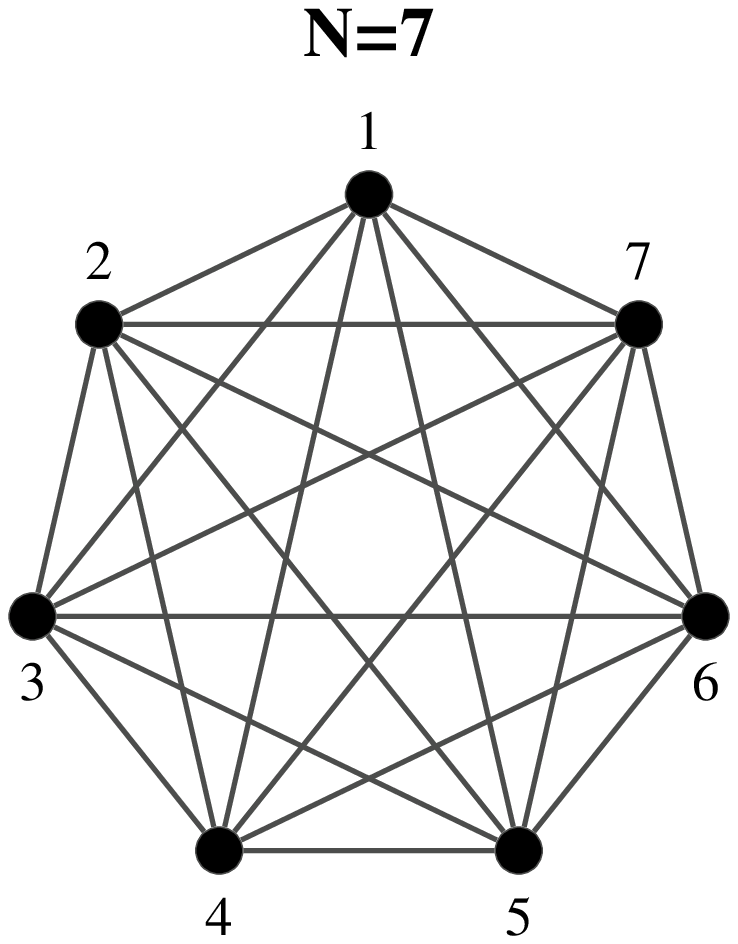} \quad \includegraphics{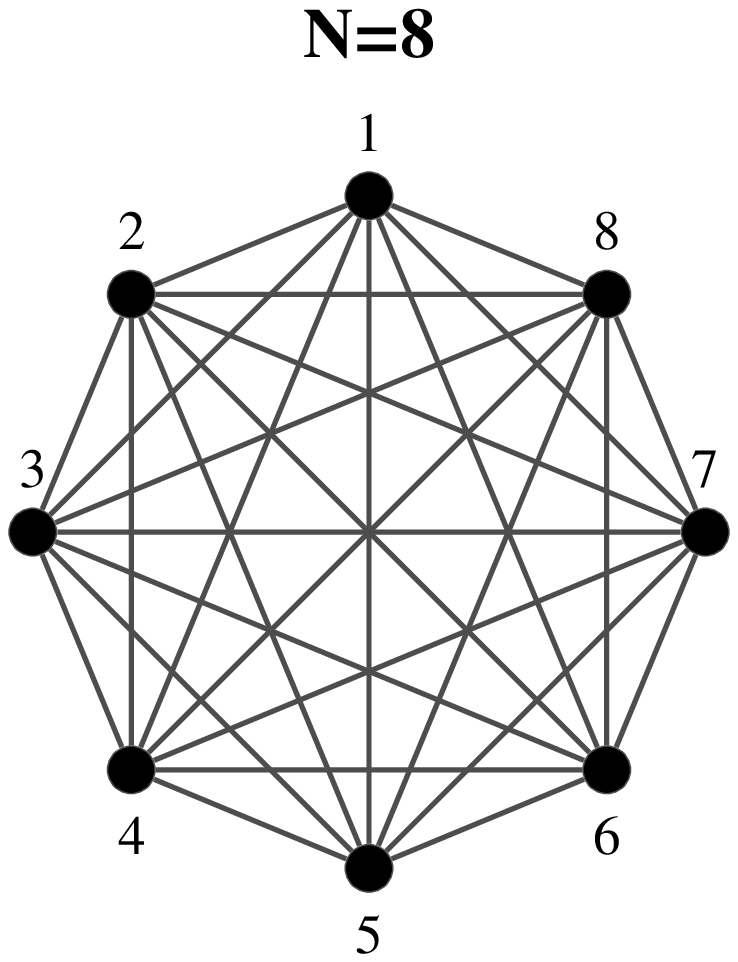}}
\caption{Diagram of the complete graph states for $N=3,4,5,6,7$ and $8$.} \label{f1}
\end{center}
\end{figure}
\subsection{Correlation tensors of multipartite entangled states} 
To begin, we find standard tensor norm of $N$-body correlation tensors of  
certain multiqubit entangled states. 
We then consider complete graph states and derive a general expression 
for the standard norm of $N$-body correlation tensors. 
We also consider GHZ, W and cluster states and evaluate the standard tensor norms 
and compare them with the complete graph states. 

Graph states are pure multiqubit entangled states conventionally 
represented by mathematical graphs \cite{hein2004}. 
In our study, we consider only complete graph (CG) states in which 
each vertex is connected with the remaining vertices through edges. 
The graph state is represented by \cite{hein2005}
\begin{align}
|G_N\rangle = \prod_{\{a,b\}\in E} U^{\{ab\}} |+\rangle^{\otimes N}. \label{graph}
\end{align}
In the above $a$ and $b$ are qubits, $E$ is the edge between $a$ and $b$ 
which in turn represents the interaction of 
two vertices or qubits, $N$ is the number of vertices or qubits, 
$|+\rangle$ is the single qubit state $\left(=\frac{|0\rangle+|1\rangle}{\sqrt{2}}\right)$ 
and $U^{\{ab\}}$ is the controlled-$Z$ operation between the qubits $a$ and $b$ which  
can be represented by the following matrix 
\begin{align}
U^{\{ab\}}=
\begin{pmatrix}
1 & 0 & 0 & 0\\
0 & 1 & 0 & 0\\
0 & 0 & 1 & 0\\
0 & 0 & 0 & -1
\end{pmatrix}. 
\end{align}
For instance, a $3$-qubit complete graph state is represented by   

\begin{align}
|G_3\rangle = \frac{1}{\sqrt{8}} \big( |000\rangle + |001\rangle +|010\rangle -|011\rangle +|100\rangle -|101\rangle - |110\rangle - |111\rangle \big).  
\end{align}
In Figure \ref{f1} we depict the complete graph states for some vertices. 
Now let us determine the norm of 
$N$-body correlation tensors for $N$-qubit complete graph states. 

The one-body correlation tensor of a single qubit is $||T_{i_1}||=1$. 
As far as the two qubit complete graph state is concerned, 
we have to evaluate nine two-body correlation tensors, that is  
$\mathcal{T}_{i_1,i_2}=\langle \sigma_{i_1}^{(1)}\otimes \sigma_{i_2}^{(2)} \rangle$, $i_1,i_2=1,2,3$. 
Evaluating all of them  
($\mathcal{T}_{1,1}=0$, $\mathcal{T}_{1,2}=0$, $\mathcal{T}_{1,3}=1$, $\mathcal{T}_{2,1}=0$, $\mathcal{T}_{2,2}
=1$, $\mathcal{T}_{2,3}=0$, $\mathcal{T}_{3,1}=1$, $\mathcal{T}_{3,2}=0$, $\mathcal{T}_{3,3}=0$) 
and substituting them into (\ref{st}),  
$||T_{i_1,i_2}|| = \sqrt{\sum_{i_1,i_2=1}^3 \mathcal{T}_{i_1,i_2}^2}$, 
we find $||T_{i_1,i_2}||=\sqrt{3}$. 
In the case of $3$-qubit complete graph state, we have to evaluate $27$ correlation tensors 
$\mathcal{T}_{i_1,i_2,i_3}=\langle \sigma_{i_1}^{(1)}\otimes \sigma_{i_2}^{(2)}\otimes \sigma_{i_3}^{(3)} \rangle$, 
$i_1,i_2,i_3=1,2,3$. Evaluating all of them and substituting the resultant values into (\ref{st}), we find  
the standard tensor norm of $3$-body correlation tensors is $||T_{i_1,i_2,i_3}||=\sqrt{4}$. 
For the $N$-qubit complete graph states we have to determine 
$3^N$ correlation tensors from which we can fix the standard tensor norm of $N$-body correlation tensors. 
In the case of $N$-qubit complete graph states, we observe that most of the correlation tensors become zero and 
in the remaining cases each correlation tensor produces the value one. 
From this outcome we identify the non-zero expectation value of 
tensor product of Pauli matrices for $N$-qubit complete graph states. 
The standard tensor norm of non-zero correlation tensors reads  
\begin{subequations}
\begin{align}
||T_{i_1,i_2,\ldots,i_N}||_{\text{odd}}^2 =& \sum_{x=2i-1, \atop i=1,2,\ldots,\frac{N+1}{2}} \left( \sum_l P_l \left\langle \sigma_1^{\otimes x} \otimes \sigma_3^{\otimes (N-x)} \right\rangle^2\right), 
\label{stnopercg1}\\
||T_{i_1,i_2,\ldots,i_N}||_{\text{even}}^2 =& \sum_{x=2i-1, \atop i=1,2,\ldots,\frac{N}{2}} \left( \sum_l P_l \left\langle \sigma_1^{\otimes x} \otimes \sigma_3^{\otimes (N-x)} \right\rangle^2 
\right) + \langle \sigma_2^{\otimes N} \rangle^2,  
\label{stnopercg2}
\end{align}
\label{stnopercg}
\end{subequations}
for odd and even $N$ respectively. Here $\sum_l P_l$ denotes the sum over all possible permutations.  
The operators given in (\ref{stnopercg}) and their possible permutations 
provide non-zero expectation value for the $N$-qubit complete graph states. 
By counting the number of permutations, we can derive a general expression 
for the norm of $N$-body correlation tensors of $N$-qubit complete graph states. 
Since each expectation value is one, the resultant value can be expressed in the form      
\begin{align}
||T_{i_1,i_2,\ldots,i_N}||_{\text{CG}} = \sqrt{2^{N-1}+s}, \label{stng}
\end{align}
where $s=0$ $(1)$, if $N$ is odd (even). 
We demonstrate the equivalence of (\ref{stnopercg}) and (\ref{stng}) in the Appendix. 

It is known that GHZ state is locally equivalent to the complete graph state \cite{hein2005}. 
The standard tensor norm of correlation tensors of $N$-qubit GHZ state,     
$|GHZ_N\rangle = \frac{1}{2}\left( |0^{\otimes N}\rangle + |1^{\otimes N}\rangle \right)$ \cite{gree1989}, 
matches with the expression given in (\ref{stng}). 
But in this case, the non-zero correlation tensors are found to be 
\begin{subequations}
\begin{align}
||T_{i_1,i_2,\ldots,i_N}^{GHZ}||_{\text{odd}}^2 =& \sum_{x=2j, \atop i=0,1,\ldots,\frac{N-1}{2}} \left( \sum_l P_l \left\langle \sigma_2^{\otimes x} \otimes \sigma_1^{\otimes (N-x)} \right\rangle^2\right), \\
||T_{i_1,i_2,\ldots,i_N}^{GHZ}||_{\text{even}}^2 =& \sum_{x=2j, \atop i=0,1,\ldots,\frac{N}{2}} \left( \sum_l P_l \left\langle \sigma_2^{\otimes x} \otimes \sigma_1^{\otimes (N-x)} \right\rangle^2 
\right) + \langle \sigma_3^{\otimes N} \rangle^2,    
\end{align}
\end{subequations}
which are different from Eq.(\ref{stnopercg}). 

In similar manner we can also determine the standard tensor norm of correlation tensors 
of $N$-qubit W states   
$(|W_N\rangle$ $= \frac{1}{\sqrt{N}} \left(|00\ldots 1\rangle + |0\ldots 10\rangle+ \ldots +|10\ldots 0\rangle \right))$ \cite{dur2000} 
and cluster states  
$(|C_N\rangle = \frac{1}{2^{N/2}} \bigotimes_{a=1}^N \left( |0\rangle_a \sigma_z^{(a+1)} + |1\rangle_a \right)$ ~~ 
with $\sigma_z^{(N+1)}=1$ \cite{brie2001} for several $N$. 
The results are given in Table \ref{stnctmes}. 

\begin{table}[t]
\caption{Standard tensor norm of $N$-body correlation tensors for multipartite entangled states}
\label{stnctmes}
\begin{center}
 \begin{tabular}{ccccc} 
\hline
\hline
$N$ &  CG & GHZ & W & Cluster\\  
\hline
\hline
$2$ & $\sqrt{3}$ & $\sqrt{3}$ & $\sqrt{3}$ & $\sqrt{3}$\\ 
$3$ & $\sqrt{4}$ & $\sqrt{4}$ & $\sqrt{\frac{11}{3}}$ &  $\sqrt{4}$\\
$4$ & $\sqrt{9}$ & $\sqrt{9}$ & $\sqrt{4}$ &   $\sqrt{5}$ \\
$5$ & $\sqrt{16}$ & $\sqrt{16}$ & $\sqrt{\frac{21}{5}}$ &  $\sqrt{8}$ \\
$6$ & $\sqrt{33}$ & $\sqrt{33}$ & $\sqrt{\frac{13}{3}}$ & $\sqrt{12}$ \\
$7$ & $\sqrt{64}$ & $\sqrt{64}$ & $\sqrt{\frac{31}{7}}$ &  $\sqrt{17}$ \\
$8$ & $\sqrt{129}$ & $\sqrt{129}$ & $\sqrt{\frac{9}{2}}$ &  $\sqrt{25}$ \\
\hline
\end{tabular}
\end{center}
\end{table}
\begin{figure}[t]
\begin{center}
\resizebox{.5\columnwidth}{!}{\includegraphics{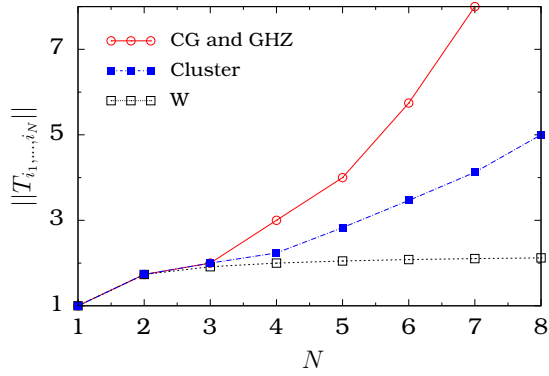}}
\caption{Norm of correlation tensors for multipartite entangled states.} \label{cormulent} 
\end{center}
\end{figure}
In Fig. \ref{cormulent}, we plot the standard tensor norm of different multipartite entangled states against $N$.   
The figure reveals that the tensor norm of complete graph state and the GHZ state coincides and 
shows rapid increase for large $N$ compared to the W and cluster states, 
which in turn provide a suitable upper bound for the $k$-separability. 
Hence, to derive $k$-separability criterion for a class of complete graph states, 
including $N$-qubit GHZ state, we use the expression (\ref{stng}) to construct the upper bound. 
%%%%%%%%%%%%%%%%%%%%%%%%%%%%%%%%%%%%%%%%%%%%%%%%%%%%%%%%%%%%%%%%%%%%%%%%%%%%%%%%%%%%%%%%%%%%%%%%%%%%%%%%%%%%%%%%%%%%%%%%%%%%%%%%%%%%%%%%%%%%%%%%%%%%%%%%%
\section{Criterion for non-$k$-separability}
\label{sec3}
In this section, we present a sufficient condition to identify the 
non-$k$-separability of a class of complete graph states. 
We derive this condition based on correlation tensors \cite{julio2011}. 
We mention here that our criterion is applicable only to 
pure complete graph states and to probabilistic mixtures of these. \\

\noindent{\bf Criterion : }
Let $\rho$ be a class of $N$-qubit complete graph state. If $\rho$ is $k$-separable, then 
\begin{align}
|| T_{i_1,i_2,\ldots,i_N}|| \leq \max_j \left\{ \prod_{i=1}^k \sqrt{2^{m_i^{(j)}-1}+s_i^{(j)}} \right\}.   
\label{cond}
\end{align}
Here for each $j$ there exists at most one $i$ such that $m_i^{(j)}=2$. 
The above condition is advantageous in situations where more than two-body correlations are relevant \cite{julio2011}. 
If the inequality (\ref{cond}) does not hold, then $\rho$ is a non-$k$-separable state. 

In Eq.(\ref{cond}), $m_i^{(j)}$'s denote the number of subsystem(s) in each compartment of $j^{\text{th}}$ 
partition that belongs to $k$-separable $N$-partite state 
and $m_1^{(j)}+m_2^{(j)}+m_3^{(j)}+\ldots+m_k^{(j)}=N$ with $1\leq m_i^{(j)} \leq N-1$.
We should consider $s_i^{(j)}=0$ whenever $m_i^{(j)}$ is odd and $s_i^{(j)}=1$ in case $m_i^{(j)}$ is even. 
Here $j=1,2,\ldots,C$ with $C$ is the total number of possible partitions.  \\ 

\noindent {\bf Proof : } 
To prove the inequality (\ref{cond}) 
we adopt the methodology given in \cite{julio2011,julio2007}. 

Let us consider a pure $k$-separable $N$-qubit state 
\begin{align}
|\psi_{k\textrm{-sep}}\rangle =& |\psi_1\rangle_{i_1,i_2,\ldots,i_{N_1}} \otimes |\psi_2\rangle_{i_{N_1+1},i_{N_1+2},\ldots,i_{N_2}} \otimes \cdots \notag\\
& \otimes|\psi_{k-1}\rangle_{i_{N_{N-2}+1},i_{N_{N-2}+2},\ldots,i_{N_{N-1}}} \otimes |\psi_{k}\rangle_{i_{N_{N-1}+1},i_{N_{N-1}+2},\ldots,i_{N_N}}, \label{app1}
\end{align} 
where $A_1=\{i_1,i_2,\ldots,i_{N_1}\}$, $A_2=\{i_{N_1+1},$ $i_{N_1+2},\ldots,i_{N_2}\}$, $\ldots$, 
$A_k=\{i_{N_{N-1}+1},i_{N_{N-1}+2},\ldots,i_{N_N}\}$, $\{ i_1, i_2, \ldots, i_{N_N}\} = \{ 1,2, \ldots, N\}$ 
and $|\psi_i\rangle$ ($1\leq i\leq k$) should be within the class of complete graph states. 
The standard tensor norm of $N$-qubit state can be factorized into product 
of norm of $k$ number of substates, that is 
\begin{align} 
|| T_{i_1,i_2,\ldots,i_{N_N}}||  =& || T_{i_1,\ldots,i_{N_1}}||~|| T_{i_{N_1+1},i_{N_1+2},\ldots,i_{N_2}}||
~\cdots~|| T_{i_{N_{N-1}+1},i_{N_{N-1}+2},\ldots,i_{N_N}}||.  
\label{tnks}
\end{align}
We determine the norm of correlation tensors of each $|\psi_i\rangle$ (vide Eq.(\ref{stng})). 
We then substitute the resultant expressions into (\ref{tnks}) with new labels $m_i$ and $s_i$. 
Doing so, we obtain 
\begin{align} 
|| T_{i_1,i_2,\ldots,i_{N_N}}||^{(j)}  =& \left(\sqrt{2^{m_1^{(j)}-1}+s_1^{(j)}}\right)\times\left(\sqrt{2^{m_2^{(j)}-1}+s_2^{(j)}}\right)\nonumber\\
& \times\cdots\times\left(\sqrt{2^{m_k^{(j)}-1}+s_k^{(j)}}\right),    
\label{tnks2}
\end{align} 
for any partition $j$. Simplifying the expression (\ref{tnks2}), we find  
\begin{align}
|| T_{i_1,i_2,\ldots,i_{N_N}}||^{(j)} = \prod_{i=1}^k \sqrt{2^{m_i^{(j)}-1}+s_i^{(j)}}.
\label{tnks23}
\end{align}
To extend the condition (\ref{tnks23}) for mixed $N$-qubit states, 
we impose the convexity property on the norm. 
Here the convexity is guaranteed by triangle inequality \cite{julio2011}. 
Hence Eq.(\ref{tnks23}) for mixed $N$-qubit state can be written as 
\begin{align}
|| T_{i_1,i_2,\ldots,i_{N_N}}|| \leq \sum_j p_j \left(\prod_{i=1}^k \sqrt{2^{m_i^{(j)}-1}+s_i^{(j)}}\right),  
\label{tnks24}
\end{align}
where $0\leq p_j \leq 1$. In Eq.(\ref{tnks24}), the probabilistic mixture of $k$-separable pure states provide an 
upper bound for $k$-separable $N$-qubit states within the class of complete graph states. 

We note that, in different partitions of $k$-separable $N$-qubit pure states, 
the state which have maximum norm can be considered as an upper bound. 
Our analysis on the norm of different partitions of $k$-separable $N$-qubit pure states 
reveal that the partition which consists of one-body correlation, more than two-body correlations 
and one two-body correlation will provide the upper bound. 
The presence of more than one two-body correlation always gives a higher value. 
Hence it does not form an optimal bound. 
For example, the norm of $3$-separable $5$-qubit state under $1|2|2$ partition is equal to 
the norm of $2$-separable $5$-qubit state under $1|4$ partition. 
So we consider only the partitions which consist of at most one two-body correlation.   
From these observations, we conclude that $k$-separable $N$-qubit states within the class of 
complete graph states hold the inequality 
\renewcommand{\theequation}{\arabic{equation}}
\begin{align}
\setcounter{equation}{10}
|| T_{i_1,i_2,\ldots,i_{N_N}}|| \leq \max_j \left\{ \prod_{i=1}^k \sqrt{2^{m_i^{(j)}-1}+s_i^{(j)}} \right\},  
\end{align}
where more than two body correlations are relevant. 
This completes the proof for the proposed criterion. 

Due to more number of partitions, 
it is very difficult to derive an explicit expression that gives an upper bound 
for the $k$-separable $N$-qubit state.  
However, a compact expression for biseparable states can be given as  
\renewcommand{\theequation}{\arabic{equation}}
\begin{align}
\setcounter{equation}{16}
|| T_{i_1,i_2,\ldots,i_N}|| \leq \sqrt{\left(2^{b-1}+s_{b}\right)~\left(2^{(N-{b})-1}+s_{(N-b)}\right)}, 
\label{bicond}
\end{align}
where $b=1$ if $\left[\left[\frac{N}{2}\right]\right]\leq 2$ otherwise $b=2$, 
where $\left[\left[\frac{N}{2}\right]\right]$ is the least integer of $\frac{N}{2}$. 
Here $s_b=0$ $(1)$, if $b$ is odd (even) and $s_{(N-b)}=0$ $(1)$, if $N-b$ is odd (even).  
The condition (\ref{bicond}) acts as a sufficient condition 
to identify genuine multipartite entanglement of a class of complete graph states. 

Further, we determine a suitable upper bound for several $k$-separable $N$-qubit states 
and present them in Table \ref{stnparti}. 
\begin{table}[h]
\caption{Upper bound values for $2$-, $3$- and $4$-separable 
$N$-qubit complete graph states}
\label{stnparti}
\begin{center}
 \begin{tabular}{crrr} 
\hline
\hline
$N$ & $2$-sep  & $3$-sep & $4$-sep \\  
\hline
\hline
$3$ & 1.7320 $(1|2)$ & 1 $(1|1|1)$ & $-$ \\ 
$4$ & 2 $(1|3)$ & 1.7320 $(1|1|2)$ & 1 $(1|1|1|1)$ \\
$5$ & 3.4641 $(2|3)$ & 2 $(1|1|3)$ & 1.7320 $(1|1|1|2)$ \\ 
$6$ & 5.1961 $(2|4)$ & 3.4640 $(1|2|3)$ & 2 $(1|1|1|3)$ \\
$7$ & 6.9282 $(2|5)$ & 5.1961 $(1|2|4)$ & 3.4641 $(1|1|2|3)$ \\
$8$ & 9.9498 $(2|6)$ & 6.9282 $(1|2|5)$ & 5.1961 $(1|1|2|4)$ \\ 
$9$ & 13.8564 $(2|7)$ & 9.9498 $(1|2|6)$ & 6.9282 $(1|1|2|5)$ \\
\hline
\end{tabular}
\end{center}
\end{table}
    
%%%%%%%%%%%%%%%%%%%%%%%%%%%%%%%%%%%%%%%%%%%%%%%%%%%%%%%%%%%%%%%%%%%%%%%%%%%%%%%%%%%%%%%%%%%%%%%%%%%%%%%%%%%%%%%%%%%%%%%%%%%%%%%%%%%%%%%%%%%%%%%%%%%%%%%%%%%%%%
\begin{figure}[h]
\begin{center}
\resizebox{.5\columnwidth}{!}{\includegraphics{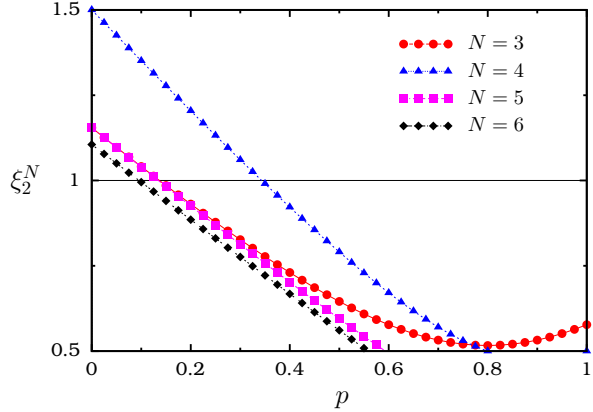}}
\caption{Genuine multipartite entanglement of $N$-qubit complete graph states added with colored product noise.} \label{9qugm}
\end{center}
\end{figure}
\section{Example} 
\label{sec4}
In this section, we study the robustness of our criterion against noise. 
To study the noise robustness of our separability criterion, 
one may consider a state mixed with white noise (identity operator), see for example Ref. \cite{guhne2009}. 
Since the correlation tensor elements vanish for the identity operator,  
adding white noise with a complete graph state is not a meaningful one.  
Therefore we add a colored product noise and 
identify the GME and non-$k$-separability of $N$-qubit complete graph states mixed 
with colored product noise \cite{arij2015}.

We consider the $N$-qubit complete graph states added with colored product noise 
\begin{align}
\rho_G = (1-p) |G_N\rangle \langle G_N| + p |1\rangle\langle 1|^{\otimes N},  
\label{rhogcpn}
\end{align}
where $|G_N\rangle$ is the $N$-qubit complete graph state (\ref{graph})  
and $p$ $(0\leq p\leq 1)$ is the probability. 
We note here that the state (\ref{rhogcpn}) is in the form to which 
our criterion can be verified unambiguously. 
 
Enforcing the condition (\ref{bicond}) on $\rho_G$, we find  
\begin{align}
\xi_2^N = \frac{(2^{N-1}+s)(1-2p)+(2^{(N-1)}+s+1)p^2}{(2^{b-1}+s_b)~(2^{(N-b)-1}+s_{(N-b)})}.  
\label{kscond}
\end{align}
When the state $\rho_G$ obeys the inequality $\xi_2^N>1$, 
then the state is genuinely $N$-qubit entangled.  
To demonstrate this, we plot the function $\xi_2^N$ against the probability $p$ $(0\leq p\leq 1)$ 
for various values of $N$ in Fig. \ref{9qugm}. 
In this figure, the range covered by $\xi_2^N>1$ reveals the genuine $N$-qubit entanglement. 
We also analyze the genuine $N$-qubit entanglement for large $N$ values of $\rho_G$. 
In this case our criterion identifies the genuine $N$-qubit entanglement 
approximately in the range of $p\leq 0.125$. 

\begin{figure}[h]
\begin{center}
\resizebox{.5\columnwidth}{!}{\includegraphics{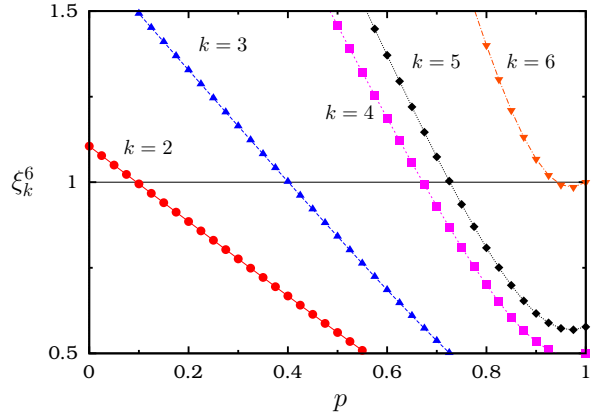}}
\caption{Non-$k$-separability of $N$-qubit complete graph states added with colored product noise.} \label{9quns} 
\end{center}
\end{figure}

To illustrate the non-$k$-separability, we consider $N=6$ in (\ref{rhogcpn}). 
We then find an upper bound for various $k$-separable $(k=2,3,4,5$ and $6)$ 6-qubit state, that is 
$\sqrt{27}$, $\sqrt{12}$, $\sqrt{4}$, $\sqrt{3}$ and $\sqrt{1}$ respectively.  
Applying the inequality (\ref{cond}) on the $6$-qubit complete graph state 
added with colored product noise, we find 
\begin{align}
\xi_k^6 = \frac{33-66~p+34~p^2}{\text{upper bound}}. 
\label{6qksepub}
\end{align}  
The case, $\xi_k^6>1$ confirms the non-$k$-separability. 
The outcome is drawn in Fig. \ref{9quns}.  
We have also analysed the non-full separability of $N$-qubit complete graph states added with colored product noise.  
The non-full separability of $\rho_G$ can be obtained from the expression 
\begin{align}
\xi_N^N = (2^{N-1}+s)(1-2p)+(2^{(N-1)}+s+1)p^2. 
\label{fullcond}
\end{align} 

Let us consider the $N$-qubit GHZ states mixed with colored product noise 
\begin{align} 
\rho_{GHZ} = (1-p) |GHZ_N\rangle\langle GHZ_N| + p |1\rangle\langle 1|^{\otimes N}, 
\end{align}
where $|GHZ_N\rangle = \frac{1}{\sqrt{2}}\left(|0\rangle^{\otimes N}+|1\rangle^{\otimes N}\right)$. 
Enforcing the condition (\ref{cond}) on $\rho_{GHZ}$, we obtain  
a function that matches with (\ref{kscond}) and (\ref{6qksepub}). 
This in turn confirms that our criterion can also detect 
non-$k$-separability in $N$-qubit complete graph states mixed with colored product noise and 
$N$-qubit GHZ state mixed with colored product noise in the same parameter range. 

%%%%%%%%%%%%%%%%%%%%%%%%%%%%%%%%%%%%%%%%%%%%%%%%%%%%%%%%%%%%%%%%%%%%%%%%%%%%%%%%%%%%%%%%%%%%%%%%%%%%%%%%%%%%%%%%%%%%%%%%%%%%%%%%%%%%%%%%%%%%%%%%%%%%%%%
\section{Conclusion}
\label{sec5}
We have proposed a sufficient condition to identify the non-$k$-separability of 
a class of mixed $N$-qubit complete graph states using correlation tensors.  
Our criterion performs well in identifying GME and non-$k$-separability of $N$-qubit complete graph states 
added with colored product noise. 
We have confirmed that our criterion also provides the same parameter range while detecting 
non-$k$-separability of the $N$-qubit complete graph states added with colored product noise 
and $N$-qubit GHZ states added with colored product noise.  
We recall here that suppose a separability criterion is derived for arbitrary states, 
one can use that criterion for any states without prior information. 
If it is not the case, one should have theoretical expectations of what the state should look like. 
In the latter case one should choose a suitable criterion 
which can detect the state under consideration \cite{julio2011}.  
Since our criterion is applicable when there is a promise that the state is a complete graph state 
(or a mixture of these), one should have theoretical expectations 
about complete graph states.      
To evaluate the standard tensor norm of our criterion 
one needs to measure totally $3^N$ correlation tensors for $N$-qubit states. 
Since a considerable number of expectation values of local observables become zero \cite{julio2011},  
one essentially performs the measurements corresponding to non-vanishing expectation values only.  
For example, in the case of $N$-qubit complete graph states added with colored product noise, 
it is enough to measure only $2^{N-1}+s+1$ (if $N$ is odd $s=0$, otherwise $s=1$) correlation tensors out of 
$3^N$ correlation tensors.  
In which, $2^{N-1}+s$ number of observables are given in equation (\ref{stnopercg}) and 
the other observable is $\langle \sigma_3^{\otimes N} \rangle$. 
Therefore $2^{N-1}+s+1$ number of local observables are sufficient 
to identify the non-$k$-separability of (\ref{rhogcpn}) through our criterion. \\

\noindent We thank the referee for his/her valuable suggestions to improve the quality of this paper. 

\section*{Author contribution statement}
Both the authors have contributed equally to the research 
and to the writing up of the paper.

\section*{Appendix}
\renewcommand{\theequation}{A.\arabic{equation}} 
To derive a general expression for the standard norm of $N$-body correlation tensors 
of $N$-qubit complete graph states, we consider Eq.(\ref{stnopercg}). 
For a given $N$, we expand the expression (\ref{stnopercg}) and find all possible 
permutations. Since each expectation value is one for $N$-qubit complete graph states, 
we just count the number of expectation values, which in turn provide the 
norm of $N$-body correlation tensors of $N$-qubit complete graph states. 
In this way we obtain the expression (\ref{stng}). 

To demonstrate the above, we consider the case $N=10$. 
Now we examine whether the expressions (\ref{stnopercg2}) 
and (\ref{stng}) yield the same value or not. 

Substituting $N=10$ in Eq. (\ref{stnopercg2}), it becomes   
\begin{align}
\setcounter{equation}{0}
||T_{i_1,i_2,\ldots,i_{10}}||_{\text{even}}^2 =& \sum_{x=1,3,5,7,9} \left( \sum_l P_l \left\langle \sigma_1^{\otimes x} \otimes \sigma_3^{\otimes (10-x)} \right\rangle^2 
\right) + \langle \sigma_2^{\otimes 10} \rangle^2.  
\end{align}
Upon expanding the right hand side, we obtain 
\begin{align}
||T_{i_1,i_2,\ldots,i_{10}}||_{\text{even}}^2 =& \sum_l P_l \left\langle \sigma_1^{\otimes 1} \otimes \sigma_3^{\otimes 9} \right\rangle^2 
 +\sum_l P_l \left\langle \sigma_3^{\otimes 1} \otimes \sigma_3^{\otimes 7} \right\rangle^2 
 +\sum_l P_l \left\langle \sigma_3^{\otimes 5} \otimes \sigma_3^{\otimes 5} \right\rangle^2 \nonumber\\
& +\sum_l P_l \left\langle \sigma_3^{\otimes 7} \otimes \sigma_3^{\otimes 3} \right\rangle^2 
 +\sum_l P_l \left\langle \sigma_3^{\otimes 9} \otimes \sigma_3^{\otimes 1} \right\rangle^2 
+ \langle \sigma_2^{\otimes 10} \rangle^2.  
\label{appeven1}
\end{align}
The total number of permutations of $a$ and $b$ in $(a^{\otimes p}\otimes b^{\otimes q})$ 
are $\frac{(a+b)!}{a!~b!}$. 
Using this in Eq.(\ref{appeven1}), we obtain 
\begin{align}
||T_{i_1,i_2,\ldots,i_{10}}||_{\text{even}}^2 =& \frac{10!}{1!~9!} + \frac{10!}{3!~7!} + \frac{10!}{5!~5!} 
 + \frac{10!}{7!~3!} + \frac{10!}{9!~1!} + 1. 
\label{appeven2}
\end{align}
Simplifying the right hand side, we find 
\begin{align}
||T_{i_1,i_2,\ldots,i_{10}}||_{\text{even}}^2 = 513.  
\label{appeven3}
\end{align}
One can obtain the same value from $\sqrt{2^{N-1}+s}$ while substituting $N=10$ and $s=1$. 

Similarly, for the case $N=21$, Eq. (\ref{stnopercg1}) yields  
\begin{align}
||T_{i_1,i_2,\ldots,i_{21}}||_{\text{odd}}^2 =& \sum_{x=1,3,5,7,9,11,\atop 13,15,17,19,21} 
\left( \sum_l P_l \left\langle \sigma_1^{\otimes x} \otimes \sigma_3^{\otimes (21-x)} \right\rangle^2 
\right).  
\end{align}
In this case, we find  
\begin{align}
||T_{i_1,i_2,\ldots,i_{21}}||_{\text{odd}}^2 =& \frac{21!}{1!~20!} + \frac{21!}{3!~18!} + \frac{21!}{5!~16!} + \frac{21!}{7!~14!} 
 + \frac{21!}{9!~12!} + \frac{21!}{11!~10!} \nonumber\\
& + \frac{21!}{13!~8!} + \frac{21!}{15!~6!} + \frac{21!}{17!~4!}  + \frac{21!}{19!~2!} + \frac{21!}{21!~0!}. 
\label{appeven22}
\end{align}
Evaluating the right hand side, we obtain $10,48,576$, 
which in turn exactly coincides with the one given in (\ref{stng}). 

%%%%%%%%%%%%%%%%%%%%%%%%%%%%%%%%%%%%
%%%%%%%%%%%%%%%%%%%%%%%%%%%%%%%%%%%%%%%%%%%%%%%%%%%%%%%%%%%%%%%%%%%%%%%%%%%%%%%%%%%%%%%%%%%%%%%%%%%%%%%%%%%%%%%%%%%

\end{document}